\documentclass{article}
\usepackage{spconfa4,amsmath,graphicx}
\newcommand{\da}{$\downarrow$}
\newcommand{\ua}{$\uparrow$}

\usepackage{graphicx} 
\usepackage[acronym]{glossaries}
\usepackage{booktabs}
\usepackage{amsmath, amssymb}
\usepackage{easyReview}
\usepackage{titlecaps}
\usepackage{enumitem}
\usepackage{xurl}
\usepackage{hyperref}
\usepackage[htt]{hyphenat}
\usepackage{pgf}
\usepackage{bm}
\usepackage{dirtytalk}

\usetikzlibrary{arrows.meta}

\makeatletter

\pgfdeclarearrow{
  name = ipe _linear,
  defaults = {
    length = +1bp,
    width  = +.666bp,
    line width = +0pt 1,
  },
  setup code = {
    \pgfarrowssetbackend{0pt}
    \pgfarrowssetvisualbackend{
      \pgfarrowlength\advance\pgf@x by-.5\pgfarrowlinewidth}
    \pgfarrowssetlineend{\pgfarrowlength}
    \ifpgfarrowreversed
      \pgfarrowssetlineend{\pgfarrowlength\advance\pgf@x by-.5\pgfarrowlinewidth}
    \fi
    \pgfarrowssettipend{\pgfarrowlength}
    \pgfarrowshullpoint{\pgfarrowlength}{0pt}
    \pgfarrowsupperhullpoint{0pt}{.5\pgfarrowwidth}
    \pgfarrowssavethe\pgfarrowlinewidth
    \pgfarrowssavethe\pgfarrowlength
    \pgfarrowssavethe\pgfarrowwidth
  },
  drawing code = {
    \pgfsetdash{}{+0pt}
    \ifdim\pgfarrowlinewidth=\pgflinewidth\else\pgfsetlinewidth{+\pgfarrowlinewidth}\fi
    \pgfpathmoveto{\pgfqpoint{0pt}{.5\pgfarrowwidth}}
    \pgfpathlineto{\pgfqpoint{\pgfarrowlength}{0pt}}
    \pgfpathlineto{\pgfqpoint{0pt}{-.5\pgfarrowwidth}}
    \pgfusepathqstroke
  },
  parameters = {
    \the\pgfarrowlinewidth,%
    \the\pgfarrowlength,%
    \the\pgfarrowwidth,%
  },
}

\pgfdeclarearrow{
  name = ipe _pointed,
  defaults = {
    length = +1bp,
    width  = +.666bp,
    inset  = +.2bp,
    line width = +0pt 1,
  },
  setup code = {
    \pgfarrowssetbackend{0pt}
    \pgfarrowssetvisualbackend{\pgfarrowinset}
    \pgfarrowssetlineend{\pgfarrowinset}
    \ifpgfarrowreversed
      \pgfarrowssetlineend{\pgfarrowlength}
    \fi
    \pgfarrowssettipend{\pgfarrowlength}
    \pgfarrowshullpoint{\pgfarrowlength}{0pt}
    \pgfarrowsupperhullpoint{0pt}{.5\pgfarrowwidth}
    \pgfarrowshullpoint{\pgfarrowinset}{0pt}
    \pgfarrowssavethe\pgfarrowinset
    \pgfarrowssavethe\pgfarrowlinewidth
    \pgfarrowssavethe\pgfarrowlength
    \pgfarrowssavethe\pgfarrowwidth
  },
  drawing code = {
    \pgfsetdash{}{+0pt}
    \ifdim\pgfarrowlinewidth=\pgflinewidth\else\pgfsetlinewidth{+\pgfarrowlinewidth}\fi
    \pgfpathmoveto{\pgfqpoint{\pgfarrowlength}{0pt}}
    \pgfpathlineto{\pgfqpoint{0pt}{.5\pgfarrowwidth}}
    \pgfpathlineto{\pgfqpoint{\pgfarrowinset}{0pt}}
    \pgfpathlineto{\pgfqpoint{0pt}{-.5\pgfarrowwidth}}
    \pgfpathclose
    \ifpgfarrowopen
      \pgfusepathqstroke
    \else
      \ifdim\pgfarrowlinewidth>0pt\pgfusepathqfillstroke\else\pgfusepathqfill\fi
    \fi
  },
  parameters = {
    \the\pgfarrowlinewidth,%
    \the\pgfarrowlength,%
    \the\pgfarrowwidth,%
    \the\pgfarrowinset,%
    \ifpgfarrowopen o\fi%
  },
}

\newdimen\ipeminipagewidth

\tikzstyle{ipe import} = [
  x=1bp, y=1bp,
  ipe node stretch/.store in=\ipenodestretch,
  ipe stretch normal/.style={ipe node stretch=1},
  ipe stretch normal,
  ipe node/.style={
    anchor=base west, inner sep=0, outer sep=0, scale=\ipenodestretch
  },
  ipe mark scale/.store in=\ipemarkscale,
  ipe mark tiny/.style={ipe mark scale=1.1},
  ipe mark small/.style={ipe mark scale=2},
  ipe mark normal/.style={ipe mark scale=3},
  ipe mark large/.style={ipe mark scale=5},
  ipe mark normal, 
  ipe circle/.pic={
    \draw[line width=0.2*\ipemarkscale]
      (0,0) circle[radius=0.5*\ipemarkscale];
    \coordinate () at (0,0);
  },
  ipe disk/.pic={
    \fill (0,0) circle[radius=0.6*\ipemarkscale];
    \coordinate () at (0,0);
  },
  ipe fdisk/.pic={
    \filldraw[line width=0.2*\ipemarkscale]
      (0,0) circle[radius=0.5*\ipemarkscale];
    \coordinate () at (0,0);
  },
  ipe box/.pic={
    \draw[line width=0.2*\ipemarkscale, line join=miter]
      (-.5*\ipemarkscale,-.5*\ipemarkscale) rectangle
      ( .5*\ipemarkscale, .5*\ipemarkscale);
    \coordinate () at (0,0);
  },
  ipe square/.pic={
    \fill
      (-.6*\ipemarkscale,-.6*\ipemarkscale) rectangle
      ( .6*\ipemarkscale, .6*\ipemarkscale);
    \coordinate () at (0,0);
  },
  ipe fsquare/.pic={
    \filldraw[line width=0.2*\ipemarkscale, line join=miter]
      (-.5*\ipemarkscale,-.5*\ipemarkscale) rectangle
      ( .5*\ipemarkscale, .5*\ipemarkscale);
    \coordinate () at (0,0);
  },
  ipe cross/.pic={
    \draw[line width=0.2*\ipemarkscale, line cap=butt]
      (-.5*\ipemarkscale,-.5*\ipemarkscale) --
      ( .5*\ipemarkscale, .5*\ipemarkscale)
      (-.5*\ipemarkscale, .5*\ipemarkscale) --
      ( .5*\ipemarkscale,-.5*\ipemarkscale);
    \coordinate () at (0,0);
  },
  /pgf/arrow keys/.cd,
  ipe arrow normal/.style={scale=1},
  ipe arrow tiny/.style={scale=.4},
  ipe arrow small/.style={scale=.7},
  ipe arrow large/.style={scale=1.4},
  ipe arrow normal,
  /tikz/.cd,
  ipe arrows/.style={
    ipe normal/.tip={
      ipe _pointed[length=1bp, width=.666bp, inset=0bp,
                   quick, ipe arrow normal]},
    ipe pointed/.tip={
      ipe _pointed[length=1bp, width=.666bp, inset=0.2bp,
                   quick, ipe arrow normal]},
    ipe linear/.tip={
      ipe _linear[length = 1bp, width=.666bp,
                  ipe arrow normal, quick]},
    ipe fnormal/.tip={ipe normal[fill=white]},
    ipe fpointed/.tip={ipe pointed[fill=white]},
    ipe double/.tip={ipe normal[] ipe normal},
    ipe fdouble/.tip={ipe fnormal[] ipe fnormal},
    ipe arc/.tip={ipe normal},
    ipe farc/.tip={ipe fnormal},
    ipe ptarc/.tip={ipe pointed},
    ipe fptarc/.tip={ipe fpointed},
  },
  ipe arrows, 
]

\tikzset{
  rgb color/.code args={#1=#2}{%
    \definecolor{tempcolor-#1}{rgb}{#2}%
    \tikzset{#1=tempcolor-#1}%
  },
}

\makeatother

\usetikzlibrary{arrows.meta,patterns}
\usetikzlibrary{ipe} 

\usepackage[style=ieee, doi=false, isbn=false, url=true, eprint=false, related=false, maxnames=5, minnames=1, date=year, block=space, citestyle=numeric-comp]{biblatex}
\AtEveryBibitem{%
    \ifboolexpr{
    test {\ifentrytype{misc}} or test {\ifentrytype{online}}
    }
    {}
    {\clearfield{url}}}
\AtEveryBibitem{\clearfield{urlyear}}
\AtEveryBibitem{\clearlist{language}}
\AtEveryBibitem{\clearfield{note}}
\AtEveryBibitem{\clearlist{location}}
\AtEveryBibitem{\clearfield{location}}
\AtEveryBibitem{\clearfield{publisher}}
\AtEveryBibitem{\clearlist{publisher}}
\AtEveryBibitem{\clearlist{pages}}
\AtEveryBibitem{\clearfield{pages}}

\everymath=\expandafter{\the\everymath\displaystyle}

\title{Revisiting Vocos: That Phasiness Business in \\ Time-Frequency Neural Vocoding}
\name{\protect\parbox{\textwidth}{\centering Ünal Ege Gaznepoğlu $^1$\sthanks{Work performed during an internship at Fraunhofer IIS.}, Frank Zalkow $^2$, Mohammad Joshaghani $^2$ \\ Emanuël A.P. Habets $^{1,2}$, Nils Peters $^3$, Christian Dittmar $^2$}}

\address{$^1$International Audio Laboratories Erlangen, Germany \\
$^2$Fraunhofer Institute for Integrated Circuits IIS, Germany \\
$^3$Dept. of Electronic \& Electrical Engineering, Trinity College Dublin, Ireland \\}
\begin{document}
\newacronym{pghi}{PGHI}{Phase Gradient Heap Integration}
\newacronym{cnn}{CNN}{convolutional neural network}
\newacronym{bpd}{BPD}{baseband phase differences}
\newacronym{fpd}{FPD}{frequency phase differences}
\newacronym{tpd}{TPD}{time phase differences}
\newacronym{lsc}{LSC}{log-spectral convergence}
\newacronym{stft}{STFT}{short-time Fourier transform}
\newacronym{mushra}{MUSHRA}{Multiple Stimuli with Hidden Reference and Anchor}
\newacronym{pqmf}{PQMF}{pseudo-quadrature mirror filters}
\newacronym{gan}{GAN}{generative adversarial networks}

\ninept
\maketitle
\begin{abstract} 
Recently, time-frequency neural vocoders have been approaching the state-of-the-art quality of time-domain neural vocoders. Vocos is a notable example due to its efficiency, but its audio quality lags behind the time-domain vocoders and the reasons remain debated. Thus, in this study, we revisit Vocos from a phase reconstruction perspective. First, we quantify the gap between time-domain and time-frequency domain vocoders using bandlimited mel spectrograms as inputs. Later, via an ablation study, we verify the Vocos architecture is effective for magnitude modeling, but less so for phase. We then adapt the Vocos backbone to predict phase differences, a precursor for phase reconstruction, and identify 1D convolutional layers are hindering their accurate prediction. Our findings indicate that future research needs to focus on inductive biases that allow the architecture to better model the time-frequency structure of speech signals, without sacrificing the support for arbitrary input representations.
\end{abstract}
\section{Introduction}
\label{sec:intro}

Neural vocoders synthesize time-domain audio signals from lossy representations, typically mel spectrograms. They play a key role in speech synthesis, such as in voice conversion and speaker anonymization systems \cite{mousaviHowShouldWe2024,gaznepogluWhyDisentanglementbasedSpeaker2025}. In these settings, time-domain neural vocoders can yield high-quality audio signals but are computationally expensive due to the cascaded temporal upsampling \cite{leeBigVGANUniversalNeural2023}. Approaches including multi-band synthesis can reduce this cost at the expense of some audio quality \cite{mustafaStreamwiseGANVocoder2021,kawamuraLightweightHighFidelityEndtoEnd2023}. Vocos \cite{siuzdakVocosClosingGap2024} avoids neural upsampling by predicting complex-valued \gls{stft} coefficients for signal reconstruction via inverse STFT. However, its audio quality is slightly below the state-of-the-art time-domain vocoders \cite{okamotoWaveNeXtConvNeXtBasedFast2023}, resulting in artifacts resembling \textit{phasiness}, the loss of clarity associated also with classical methods \cite{larochePhasevocoderThisPhasiness1997}.

Subsequent studies have examined Vocos, reaching differing interpretations and proposing divergent remedies. The original author of Vocos reports\footnote{https://openreview.net/forum?id=vY9nzQmQBw} that scaling the model does not improve the performance, and hypothesizes that since phase recovery is an autoregressive task, the \gls{cnn}-based architecture is not suitable for it \cite{morrisonChunkedAutoregressiveGAN2022}. Notably, the magnitude and phase predictions of Vocos were found to deviate substantially from the ground truth \cite{okamotoWaveNeXtConvNeXtBasedFast2023, yoneyamaWavehaxAliasingFreeNeural2025}. The inverse STFT partially compensates these errors through overlap–add redundancy, similar to the classical Griffin–Lim algorithm \cite{griffinSignalEstimationModified1984}. Few works report performance gains by incorporating loss terms between predicted and ground-truth magnitude and phase spectrograms \cite{lvFreeVFreeLunch2024,duGANNecessaryMelSpectrogramBased2025,liLearningNeuralVocoder2025} and using a pseudo-inverse for recovering the linear-scale magnitude spectrograms \cite{liLearningNeuralVocoder2025}. In WaveNeXt, the authors argue the inverse STFT is unnecessary because ConvNeXt blocks could directly predict the time-domain signal \cite{okamotoWaveNeXtConvNeXtBasedFast2023}. In WaveHax, the architecture is in the spotlight: ConvNeXt blocks with 2D convolutions refine a harmonic prior to exploit local time-frequency structure \cite{yoneyamaWavehaxAliasingFreeNeural2025}. In summary, the exact reasons for Vocos' phase modeling issues and the best way to address them remain unclear. Moreover, the aforementioned studies have not compared Vocos to time-domain vocoders on an equal footing, which prevents a clear assessment. So, in this work, we investigate why phase modeling is a bottleneck for time-frequency domain vocoders. We do so by comparing Vocos to the state-of-the-art time-domain vocoders, and later by analyzing Vocos through a phase reconstruction lens. This work has the following contributions:
\setlist[itemize]{leftmargin=1em}
\begin{itemize}
    \item We compare Vocos to BigVGAN and find that the audio quality gap is still present, even after controlling for confounding factors such as training loss and discriminators. 
    \item Training Vocos variants with oracle knowledge of either the magnitude or phase spectrograms reveals that 1D convolutions are effective for magnitude modeling, but less so for phase modeling.
    \item Then, we investigate whether the Vocos architecture can predict phase differences, a precursor for phase reconstruction that does not require autoregression, and find that it is not effective for this task either. We later show that switching to 2D convolutions substantially improves performance on this particular subtask.
\end{itemize}

\section{Methodology (Vocoding)} \label{sec:methodology}

\subsection{Input representation} \label{ssec:input_representation}
We use 80-band log mel spectrograms (for the frequency range 0--8\,kHz) as input, at a sampling rate of 22.05\,kHz. Bandlimited spectrograms could accentuate the gap between time-domain and time-frequency domain vocoders, since time-frequency domain vocoders are known to struggle with generating harmonic structure \cite{yoneyamaWavehaxAliasingFreeNeural2025}.

\subsection{Model architectures}

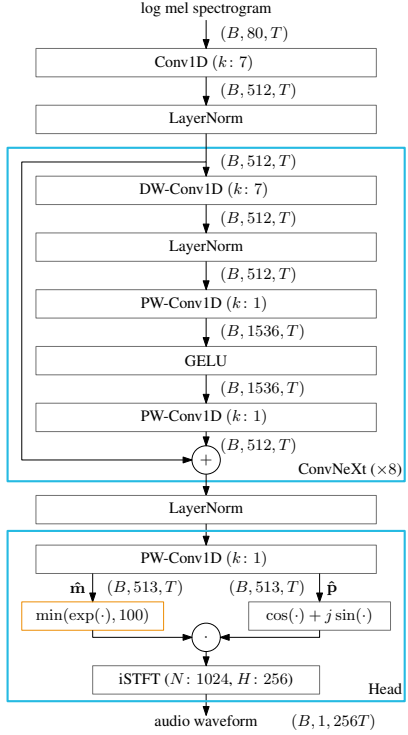
\begin{figure}[!t]
    \centering
    \resizebox{0.62\linewidth}{!}{\begingroup
\renewcommand{\baselinestretch}{1} \endlinechar=-1 \tikzstyle{ipe stylesheet} = [
  ipe import,
  even odd rule,
  line join=round,
  line cap=butt,
  ipe pen normal/.style={line width=0.4},
  ipe pen heavier/.style={line width=0.8},
  ipe pen fat/.style={line width=1.2},
  ipe pen ultrafat/.style={line width=2},
  ipe pen normal,
  ipe mark normal/.style={ipe mark scale=3},
  ipe mark large/.style={ipe mark scale=5},
  ipe mark small/.style={ipe mark scale=2},
  ipe mark tiny/.style={ipe mark scale=1.1},
  ipe mark normal,
  /pgf/arrow keys/.cd,
  ipe arrow normal/.style={scale=7},
  ipe arrow large/.style={scale=10},
  ipe arrow small/.style={scale=5},
  ipe arrow tiny/.style={scale=3},
  ipe arrow normal,
  /tikz/.cd,
  ipe arrows, 
  <->/.tip = ipe normal,
  ipe dash normal/.style={dash pattern=},
  ipe dash dotted/.style={dash pattern=on 1bp off 3bp},
  ipe dash dashed/.style={dash pattern=on 4bp off 4bp},
  ipe dash dash dotted/.style={dash pattern=on 4bp off 2bp on 1bp off 2bp},
  ipe dash dash dot dotted/.style={dash pattern=on 4bp off 2bp on 1bp off 2bp on 1bp off 2bp},
  ipe dash normal,
  ipe node/.append style={font=\normalsize},
  ipe stretch normal/.style={ipe node stretch=1},
  ipe stretch normal,
  ipe opacity 10/.style={opacity=0.1},
  ipe opacity 30/.style={opacity=0.3},
  ipe opacity 50/.style={opacity=0.5},
  ipe opacity 75/.style={opacity=0.75},
  ipe opacity opaque/.style={opacity=1},
  ipe opacity opaque,
]
\definecolor{red}{rgb}{1,0,0}
\definecolor{blue}{rgb}{0,0,1}
\definecolor{green}{rgb}{0,1,0}
\definecolor{yellow}{rgb}{1,1,0}
\definecolor{orange}{rgb}{1,0.647,0}
\definecolor{gold}{rgb}{1,0.843,0}
\definecolor{purple}{rgb}{0.627,0.125,0.941}
\definecolor{gray}{rgb}{0.745,0.745,0.745}
\definecolor{brown}{rgb}{0.647,0.165,0.165}
\definecolor{navy}{rgb}{0,0,0.502}
\definecolor{pink}{rgb}{1,0.753,0.796}
\definecolor{seagreen}{rgb}{0.18,0.545,0.341}
\definecolor{turquoise}{rgb}{0.251,0.878,0.816}
\definecolor{violet}{rgb}{0.933,0.51,0.933}
\definecolor{darkblue}{rgb}{0,0,0.545}
\definecolor{darkcyan}{rgb}{0,0.545,0.545}
\definecolor{darkgray}{rgb}{0.663,0.663,0.663}
\definecolor{darkgreen}{rgb}{0,0.392,0}
\definecolor{darkmagenta}{rgb}{0.545,0,0.545}
\definecolor{darkorange}{rgb}{1,0.549,0}
\definecolor{darkred}{rgb}{0.545,0,0}
\definecolor{lightblue}{rgb}{0.678,0.847,0.902}
\definecolor{lightcyan}{rgb}{0.878,1,1}
\definecolor{lightgray}{rgb}{0.827,0.827,0.827}
\definecolor{lightgreen}{rgb}{0.565,0.933,0.565}
\definecolor{lightyellow}{rgb}{1,1,0.878}
\definecolor{FhGblue}{rgb}{0.145,0.729,0.886}
\definecolor{FhGdarkblue}{rgb}{0,0.431,0.573}
\definecolor{FhGgray}{rgb}{0.659,0.686,0.686}
\definecolor{FhGgreen}{rgb}{0.09,0.612,0.49}
\definecolor{FhGlightblue}{rgb}{0.784,0.863,1}
\definecolor{FhGolive}{rgb}{0.694,0.784,0}
\definecolor{FhGorange}{rgb}{0.922,0.416,0.039}
\definecolor{algreen}{rgb}{0.435,0.851,0}
\definecolor{algrey}{rgb}{0.439,0.439,0.439}
\definecolor{algreyshade}{rgb}{0.851,0.851,0.851}
\definecolor{almagenta}{rgb}{0.824,0.333,1}
\definecolor{alorange}{rgb}{0.929,0.553,0.02}
\definecolor{alorangeshade}{rgb}{0.973,0.871,0.741}
\definecolor{alturquise}{rgb}{0.169,0.682,0.357}
\definecolor{alturquisedark}{rgb}{0,0.592,0.643}
\definecolor{iisgreen}{rgb}{0.09,0.612,0.49}
\definecolor{black}{rgb}{0,0,0}
\definecolor{white}{rgb}{1,1,1}
\begin{tikzpicture}[ipe stylesheet]
  \draw[shift={(48, 728)}, xscale=1.75, yscale=5.875, FhGblue, ipe pen fat]
    (0, 0) rectangle (128, -32);
  \draw[shift={(64, 784)}, xscale=1.5, yscale=0.5, algrey]
    (0, 0) rectangle (128, -32);
  \node[ipe node, anchor=center]
     at (160, 776) {Conv1D $ (k\colon 7) $};
  \draw[shift={(64, 752)}, xscale=1.5, yscale=0.5, algrey]
    (0, 0) rectangle (128, -32);
  \node[ipe node, anchor=center]
     at (160, 744) {LayerNorm};
  \draw[shift={(64, 712)}, xscale=1.5, yscale=0.5, algrey]
    (0, 0) rectangle (128, -32);
  \node[ipe node, anchor=center]
     at (160, 704) {DW-Conv1D $ (k\colon 7) $};
  \draw[-{>[ipe arrow small]}]
    (160, 768)
     -- (160, 752);
  \draw[-{>[ipe arrow small]}]
    (160, 800)
     -- (160, 784);
  \node[ipe node, anchor=west]
     at (168, 792) {$(B, 80, T)$};
  \node[ipe node, anchor=west]
     at (168, 760) {$(B, 512, T)$};
  \draw[shift={(64, 680)}, xscale=1.5, yscale=0.5, algrey]
    (0, 0) rectangle (128, -32);
  \node[ipe node, anchor=center]
     at (160, 672) {LayerNorm};
  \node[ipe node, anchor=west]
     at (168, 720) {$(B, 512, T)$};
  \node[ipe node, anchor=west]
     at (168, 688) {$(B, 512, T)$};
  \draw[-{>[ipe arrow small]}]
    (160, 696)
     -- (160, 680);
  \node[ipe node, anchor=base east]
     at (272, 544) {ConvNeXt ($\times8$)\,};
  \draw[shift={(160, 736)}, yscale=0.5]
    (0, 0)
     -- (0, -16);
  \draw[-{>[ipe arrow small]}]
    (160, 728)
     -- (160, 720)
     -- (56, 720)
     -- (56, 552)
     -- (152, 552);
  \draw[shift={(64, 648)}, xscale=1.5, yscale=0.5, algrey]
    (0, 0) rectangle (128, -32);
  \node[ipe node, anchor=center]
     at (160, 640) {PW-Conv1D $ (k\colon 1) $};
  \draw[-{>[ipe arrow small]}]
    (160, 664)
     -- (160, 648);
  \draw[shift={(64, 616)}, xscale=1.5, yscale=0.5, algrey]
    (0, 0) rectangle (128, -32);
  \node[ipe node, anchor=center]
     at (160, 608) {GELU};
  \draw[shift={(64, 584)}, xscale=1.5, yscale=0.5, algrey]
    (0, 0) rectangle (128, -32);
  \node[ipe node, anchor=center]
     at (160, 576) {PW-Conv1D $ (k\colon 1) $};
  \node[ipe node, anchor=west]
     at (168, 656) {$(B, 512, T)$};
  \node[ipe node, anchor=west]
     at (168, 624) {$(B, 1536, T)$};
  \node[ipe node, anchor=west]
     at (168, 592) {$(B, 1536, T)$};
  \node[ipe node, anchor=west]
     at (168, 560) {$(B, 512, T)$};
  \draw[-{>[ipe arrow small]}]
    (160, 632)
     -- (160, 616);
  \draw[-{>[ipe arrow small]}]
    (160, 600)
     -- (160, 584);
  \draw[-{>[ipe arrow small]}]
    (160, 568)
     -- (160, 560);
  \node[ipe node, anchor=center]
     at (160, 552) {$+$};
  \draw
    (160, 552) circle[radius=8];
  \draw[shift={(160, 544)}, yscale=0.75, -{>[ipe arrow small]}]
    (0, 0)
     -- (0, -16);
  \draw[-{>[ipe arrow small]}]
    (160, 720)
     -- (160, 712);
  \draw[shift={(64, 504)}, xscale=1.5, yscale=0.5, algrey]
    (0, 0) rectangle (128, -32);
  \node[ipe node, anchor=center]
     at (160, 496) {PW-Conv1D $ (k\colon 1) $};
  \node[ipe node, anchor=west]
     at (104, 480) {$(B, 513, T)$};
  \draw[-{>[ipe arrow small]}]
    (96, 488)
     -- (96, 472);
  \draw[-{>[ipe arrow small]}]
    (224, 488)
     -- (224, 472);
  \node[ipe node, anchor=center]
     at (96, 464) {$\min (\exp(\cdot), 100)$};
  \node[ipe node, anchor=center]
     at (224, 464) {$\cos(\cdot) + j\sin(\cdot)$};
  \draw[alorange]
    (56, 472) rectangle (136, 456);
  \draw[algrey]
    (184, 472) rectangle (264, 456);
  \node[ipe node, anchor=east]
     at (216, 480) {$(B, 513, T)$};
  \node[ipe node, anchor=center]
     at (160, 452) {$\cdot$};
  \draw
    (160, 452) circle[radius=8];
  \draw[-{>[ipe arrow small]}]
    (96, 456)
     -- (96, 452)
     -- (152, 452);
  \draw[shift={(224, 456)}, yscale=0.5, -{>[ipe arrow small]}]
    (0, 0)
     -- (0, -8)
     -- (-56, -8);
  \node[ipe node, anchor=center]
     at (160, 428) {iSTFT ($N\colon 1024$, $H\colon 256$)};
  \draw[shift={(96, 436)}, xscale=1.6, algrey]
    (0, 0) rectangle (80, -16);
  \draw[-{>[ipe arrow small]}]
    (160, 444)
     -- (160, 436);
  \draw[-{>[ipe arrow small]}]
    (160, 420)
     -- (160, 408);
  \draw[shift={(64, 532)}, xscale=1.5, yscale=0.5, algrey]
    (0, 0) rectangle (128, -32);
  \node[ipe node, anchor=center]
     at (160, 524) {LayerNorm};
  \draw[shift={(160, 516)}, yscale=0.75, -{>[ipe arrow small]}]
    (0, 0)
     -- (0, -16);
  \node[ipe node, anchor=west]
     at (208, 404) {$(B, 1, 256T)$};
  \node[ipe node, anchor=base]
     at (160, 804) {log mel spectrogram};
  \node[ipe node, anchor=north]
     at (160, 408) {audio waveform};
  \node[ipe node, anchor=east]
     at (92, 480) {$\mathbf{\hat{m}}$};
  \node[ipe node, anchor=west]
     at (228, 480) {$\mathbf{\hat{p}}$};
  \draw[FhGblue, ipe pen fat]
    (48, 511.9999) rectangle (272, 415.9999);
  \node[ipe node, anchor=base east]
     at (272, 420) {Head\,};
\end{tikzpicture}\endgroup \renewcommand{\baselinestretch}{1.5}}
    \caption{The Vocos architecture, predicting log-magnitude and phase spectrograms ($\mathbf{\hat{m}}$, $\mathbf{\hat{p}}$). DW/PW denote depthwise/pointwise operations. Batch size and number of time frames are indicated by $B$, $T$.} \vspace{-1em}
    \label{fig:vocos_arch}
\end{figure}

\begin{figure}[!t]
    \centering
    {\begingroup
\renewcommand{\baselinestretch}{1} \endlinechar=-1 \input{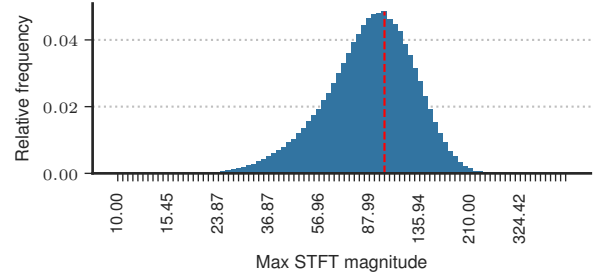}\endgroup \renewcommand{\baselinestretch}{1.5}\vspace{-1em}}
    \caption{Histogram of the maximum ground-truth STFT magnitude for all training samples (with $-1$ dBFS amplitude scaling). Dashed line indicates the Vocos' default clamping threshold of 100.} \vspace{-1em}
    \label{fig:max_stft_mag_histogram}
\end{figure}

\noindent \textbf{Vocos.} Fig.\,\ref{fig:vocos_arch} outlines the Vocos architecture. It consists of a Conv1D layer, followed by a stack of ConvNeXt \cite{liuConvNet2020s2022} blocks, and a `Head' block that interprets the network outputs as STFT coefficients. Finally, the time-domain signal is reconstructed using the inverse STFT with a Hann window. In practice, we found that the magnitude clamping (shown in orange) prevents the network from predicting accurate magnitude spectrograms. Fig.\,\ref{fig:max_stft_mag_histogram} shows the histogram of the maximum STFT magnitude for all training samples, with $-1$\,dBFS amplitude scaling. The default threshold of 100 is set too low to accurately represent up to 41\,\% of the training samples. Based on the histogram, we adopt a threshold of 400, and advise future work to carefully consider the choice of this hyperparameter. 

\vspace{0.75em}

\noindent \textbf{BigVGANv2.} We retrained BigVGAN to serve as a benchmark time-domain vocoder \cite{leeBigVGANUniversalNeural2023}. An official checkpoint \texttt{bigvgan\_v2\_22khz\_80band\_fmax8k\_256x} matches our input representation but was trained on a larger dataset, so we report results for both that checkpoint and our retrained model. We also retrained a smaller variant of BigVGANv2, with 14.4\,M parameters (denominated BigVGANv2-base), to match the parameter count of Vocos. 

\subsection{Discriminator} In our vocoder trainings, we utilize the multi-period discriminator with periods of (2, 3, 5, 7, 11) \cite{kumarHighFidelityAudioCompression2023}, EnCodec MS-STFT discriminator \cite{defossezHighFidelityNeural2023} for FFT sizes of (128, 256, 512, 1024, 2048), and the MS-SB-CQT discriminator \cite{guMultiScaleSubBandConstantQ2024}.

\section{Methodology (Phase Reconstruction)} \label{sec:methodology_phase}

In Vocos, the Head is the only block that imposes phase-related inductive bias, by (elementwise) wrapping the predicted values to the $[-\pi, \pi)$ range. However, spectrograms are characterized by a much stronger inductive bias of consistency, in the form of a coupling between the time and frequency gradients of log-magnitude and phase spectrograms \cite{augerPhaseMagnitudeRelationshipsShortTime2012,prusaNoniterativeMethodReconstruction2017}. So, \gls{stft} bins are influenced by neighbors in the time-frequency plane, and in the case of harmonic overtones, even by components spaced further apart along the frequency axis.

These relations are baked in DSP-based phase reconstruction methods, such as the \gls{pghi} \cite{prusaNoniterativeMethodReconstruction2017} as well as in some neural network-based approaches \cite{marafiotiAdversarialGenerationTimeFrequency2019,digiorgiMelSpectrogramInversion2022,masuyamaOnlinePhaseReconstruction2023,fernandezEfficientNeuralNumerical2025}. In essence, these methods first predict the phase differences (i.e., phase gradients) from a given log-magnitude spectrogram, and later integrate them across time and frequency to obtain a phase estimate. As mentioned above, Vocos does none of these operations explicitly, but instead offloads them to the ConvNeXt blocks. This warrants an investigation into whether the Vocos architecture can predict phase differences, and if not, which components are responsible for this failure. Hence, later on, we adapt the Vocos architecture to predict phase differences, and train it with a loss that directly compares the predicted and ground-truth phase differences.

\subsection{Phase differences}

Phase reconstruction methods often use the \gls{fpd} (also called group delay), \gls{tpd} (also called instantaneous frequency), and \gls{bpd}, which are given by

\begin{equation} \label{eq:fpd}
    \text{FPD}[m, n] = \mathcal{W}(\phi[m, n] - \phi[m-1, n])~,
\end{equation}
\begin{equation} \label{eq:tpd}
    \text{TPD}[m, n] = \mathcal{W}(\phi[m, n] - \phi[m, n-1])~,
\end{equation}
\begin{equation} \label{eq:bpd}
    \text{BPD}[m, n] = \mathcal{W}(\text{TPD}[m, n] - 2 \pi \frac{mH}{N})~,
\end{equation}
where $\mathcal{W}(\cdot)$ denotes the principal value wrapping operation, and $\phi$ denotes the phase spectrogram, for frequency bin $m$ and time index $n$. The STFT hop size and FFT size are denoted by $H$ and $N$, respectively. It is commonly assumed that \Gls{bpd} are easier to model with \glspl{cnn} than \gls{tpd} since cumulative contributions from the linear phase term (due to time shifts between frames) are removed \cite{krawczykSTFTPhaseReconstruction2014}. 

\subsection{Modeling and integrating phase differences}

\begin{figure}[!t]
    \centering
    \rotatebox{-90}{\resizebox{1.05\linewidth}{!}{\begingroup
\renewcommand{\baselinestretch}{1} \endlinechar=-1 \tikzstyle{ipe stylesheet} = [
  ipe import,
  even odd rule,
  line join=round,
  line cap=butt,
  ipe pen normal/.style={line width=0.4},
  ipe pen heavier/.style={line width=0.8},
  ipe pen fat/.style={line width=1.2},
  ipe pen ultrafat/.style={line width=2},
  ipe pen normal,
  ipe mark normal/.style={ipe mark scale=3},
  ipe mark large/.style={ipe mark scale=5},
  ipe mark small/.style={ipe mark scale=2},
  ipe mark tiny/.style={ipe mark scale=1.1},
  ipe mark normal,
  /pgf/arrow keys/.cd,
  ipe arrow normal/.style={scale=7},
  ipe arrow large/.style={scale=10},
  ipe arrow small/.style={scale=5},
  ipe arrow tiny/.style={scale=3},
  ipe arrow normal,
  /tikz/.cd,
  ipe arrows, 
  <->/.tip = ipe normal,
  ipe dash normal/.style={dash pattern=},
  ipe dash dotted/.style={dash pattern=on 1bp off 3bp},
  ipe dash dashed/.style={dash pattern=on 4bp off 4bp},
  ipe dash dash dotted/.style={dash pattern=on 4bp off 2bp on 1bp off 2bp},
  ipe dash dash dot dotted/.style={dash pattern=on 4bp off 2bp on 1bp off 2bp on 1bp off 2bp},
  ipe dash normal,
  ipe node/.append style={font=\normalsize},
  ipe stretch normal/.style={ipe node stretch=1},
  ipe stretch normal,
  ipe opacity 10/.style={opacity=0.1},
  ipe opacity 30/.style={opacity=0.3},
  ipe opacity 50/.style={opacity=0.5},
  ipe opacity 75/.style={opacity=0.75},
  ipe opacity opaque/.style={opacity=1},
  ipe opacity opaque,
]
\definecolor{red}{rgb}{1,0,0}
\definecolor{blue}{rgb}{0,0,1}
\definecolor{green}{rgb}{0,1,0}
\definecolor{yellow}{rgb}{1,1,0}
\definecolor{orange}{rgb}{1,0.647,0}
\definecolor{gold}{rgb}{1,0.843,0}
\definecolor{purple}{rgb}{0.627,0.125,0.941}
\definecolor{gray}{rgb}{0.745,0.745,0.745}
\definecolor{brown}{rgb}{0.647,0.165,0.165}
\definecolor{navy}{rgb}{0,0,0.502}
\definecolor{pink}{rgb}{1,0.753,0.796}
\definecolor{seagreen}{rgb}{0.18,0.545,0.341}
\definecolor{turquoise}{rgb}{0.251,0.878,0.816}
\definecolor{violet}{rgb}{0.933,0.51,0.933}
\definecolor{darkblue}{rgb}{0,0,0.545}
\definecolor{darkcyan}{rgb}{0,0.545,0.545}
\definecolor{darkgray}{rgb}{0.663,0.663,0.663}
\definecolor{darkgreen}{rgb}{0,0.392,0}
\definecolor{darkmagenta}{rgb}{0.545,0,0.545}
\definecolor{darkorange}{rgb}{1,0.549,0}
\definecolor{darkred}{rgb}{0.545,0,0}
\definecolor{lightblue}{rgb}{0.678,0.847,0.902}
\definecolor{lightcyan}{rgb}{0.878,1,1}
\definecolor{lightgray}{rgb}{0.827,0.827,0.827}
\definecolor{lightgreen}{rgb}{0.565,0.933,0.565}
\definecolor{lightyellow}{rgb}{1,1,0.878}
\definecolor{FhGblue}{rgb}{0.145,0.729,0.886}
\definecolor{FhGdarkblue}{rgb}{0,0.431,0.573}
\definecolor{FhGgray}{rgb}{0.659,0.686,0.686}
\definecolor{FhGgreen}{rgb}{0.09,0.612,0.49}
\definecolor{FhGlightblue}{rgb}{0.784,0.863,1}
\definecolor{FhGolive}{rgb}{0.694,0.784,0}
\definecolor{FhGorange}{rgb}{0.922,0.416,0.039}
\definecolor{algreen}{rgb}{0.435,0.851,0}
\definecolor{algrey}{rgb}{0.439,0.439,0.439}
\definecolor{algreyshade}{rgb}{0.851,0.851,0.851}
\definecolor{almagenta}{rgb}{0.824,0.333,1}
\definecolor{alorange}{rgb}{0.929,0.553,0.02}
\definecolor{alorangeshade}{rgb}{0.973,0.871,0.741}
\definecolor{alturquise}{rgb}{0.169,0.682,0.357}
\definecolor{alturquisedark}{rgb}{0,0.592,0.643}
\definecolor{iisgreen}{rgb}{0.09,0.612,0.49}
\definecolor{black}{rgb}{0,0,0}
\definecolor{white}{rgb}{1,1,1}
\begin{tikzpicture}[ipe stylesheet]
  \draw[shift={(72.172, 599.972)}, rotate=90, xscale=1.5, yscale=0.5, color=algrey]
    (0, 0) rectangle (128, -32);
  \node[ipe node, rotate=90, anchor=center]
     at (80.172, 695.972) {Mean subtraction};
  \draw[shift={(104.172, 599.972)}, rotate=90, xscale=1.5, yscale=0.5, color=algrey]
    (0, 0) rectangle (128, -32);
  \node[ipe node, rotate=90, anchor=center]
     at (112.172, 695.972) {FreqConv $ (k\colon 1) $};
  \draw[shift={(152.172, 591.972)}, rotate=90, xscale=0.6876, yscale=0.5]
    (0, 0) rectangle (128, -32);
  \node[ipe node, rotate=90, anchor=center]
     at (160.172, 635.98) {FreqConv $ (k\colon 1\times3) $};
  \draw[shift={(88.172, 695.972)}, rotate=90, -{>[ipe arrow small]}]
    (0, 0)
     -- (0, -16);
  \draw[shift={(56.172, 695.972)}, rotate=90, -{>[ipe arrow small]}]
    (0, 0)
     -- (0, -16);
  \node[ipe node, rotate=90, anchor=west]
     at (64.172, 703.972) {$(B, 1, F, T)$};
  \node[ipe node, rotate=90, anchor=east]
     at (176.172, 755.972) {$(B, 64, F, T)$};
  \node[ipe node, rotate=90, anchor=base]
     at (52.172, 695.972) {log mag spectrogram};
  \node[ipe node, rotate=90, anchor=west]
     at (96.172, 703.972) {$(B, 4, F, T)$};
  \node[ipe node, rotate=90, anchor=west]
     at (136.172, 699.972) {$(B, 64, F, T)$};
  \draw[shift={(128.019, 811.968)}, xscale=1.3902, yscale=1.0741, FhGblue, ipe pen fat]
    (0, 0) rectangle (72, -216);
  \draw[shift={(152.172, 715.972)}, rotate=90, xscale=0.6876, yscale=0.5, algrey]
    (0, 0) rectangle (128, -32);
  \node[ipe node, rotate=90, anchor=center]
     at (160.172, 759.98) {FreqConv $ (k\colon 1\times3) $};
  \draw[-{>[ipe arrow small]}]
    (168.172, 635.972)
     -- (216, 636)
     -- (216, 688);
  \draw[shift={(168.172, 759.972)}, rotate=90, -{>[ipe arrow small]}]
    (0, 0)
     -- (0, -16);
  \node[ipe node, rotate=90, anchor=west]
     at (176.172, 639.972) {$(B, 64, F, T)$};
  \draw[-{>[ipe arrow small]}]
    (120, 696)
     -- (144, 696)
     -- (144, 760)
     -- (152, 760);
  \draw[-{>[ipe arrow small]}]
    (144, 696)
     -- (144, 636)
     -- (152, 636);
  \draw[shift={(184.172, 715.972)}, rotate=90, xscale=0.6876, yscale=0.5, algrey]
    (0, 0) rectangle (128, -32);
  \node[ipe node, rotate=90, anchor=center]
     at (192.172, 759.98) {Sigmoid};
  \node[ipe node, rotate=90, anchor=east]
     at (208.172, 755.972) {$(B, 64, F, T)$};
  \node[ipe node, rotate=90, anchor=center]
     at (216.172, 695.972) {$\cdot$};
  \draw
    (216.1742, 695.9715) ellipse[x radius=7.9967, y radius=8.0203, rotate=90];
  \node[ipe node, rotate=90, anchor=base]
     at (384.17, 695.972) {BPD or FPD};
  \draw[-{>[ipe arrow small]}]
    (200, 760)
     -- (216, 760)
     -- (216, 704);
  \node[ipe node, rotate=90, anchor=base east]
     at (224.172, 807.98) {FreqGatedConv\,};
  \draw[shift={(240.172, 599.972)}, rotate=90, xscale=1.5, yscale=0.5, FhGblue, ipe pen fat]
    (0, 0) rectangle (128, -32);
  \node[ipe node, rotate=90, anchor=center]
     at (248.17, 695.972) {FreqGatedConv};
  \node[ipe node, rotate=90, anchor=center]
     at (272.172, 695.972) {$+$};
  \draw
    (272.172, 695.972) circle[radius=8];
  \draw[shift={(224, 696)}, xscale=1.3333, -{>[ipe arrow small]}]
    (0, 0)
     -- (12, 0);
  \draw[-{>[ipe arrow small]}]
    (232, 696)
     -- (232, 800)
     -- (272, 800)
     -- (272, 704);
  \draw[shift={(256, 696)}, xscale=0.6667, -{>[ipe arrow small]}]
    (0, 0)
     -- (12, 0);
  \draw[shift={(296.172, 599.972)}, rotate=90, xscale=1.5, yscale=0.5, FhGblue, ipe pen fat]
    (0, 0) rectangle (128, -32);
  \node[ipe node, rotate=90, anchor=center]
     at (304.17, 695.972) {FreqGatedConv};
  \node[ipe node, rotate=90, anchor=center]
     at (328.172, 695.972) {$+$};
  \draw[shift={(280, 696)}, xscale=1.0035, -{>[ipe arrow small]}]
    (0, 0)
     -- (16.172, -0.028);
  \draw[shift={(312.17, 695.972)}, rotate=90, yscale=0.4946, -{>[ipe arrow small]}]
    (0, 0)
     -- (0, -16);
  \node[ipe node, rotate=90, anchor=west]
     at (280.17, 703.972) {$(B, 64, F, T)$};
  \node[ipe node, rotate=90, anchor=west]
     at (320.17, 703.972) {$(B, 64, F, T)$};
  \draw
    (328.172, 695.972) circle[radius=8];
  \draw[shift={(344.172, 599.972)}, rotate=90, xscale=1.5, yscale=0.5, algrey]
    (0, 0) rectangle (128, -32);
  \node[ipe node, rotate=90, anchor=center]
     at (352.17, 695.972) {FreqConv $ (k\colon 1) $};
  \draw[-{>[ipe arrow small]}]
    (336, 696)
     -- (344, 696);
  \node[ipe node, rotate=90, anchor=west]
     at (368.17, 703.972) {$(B, 1, F, T)$};
  \draw[shift={(360.17, 695.972)}, rotate=90, yscale=0.7473, -{>[ipe arrow small]}]
    (0, 0)
     -- (0, -16);
  \node[ipe node, rotate=90, anchor=base east]
     at (252.172, 791.98) {$\times 2$\,};
  \node[ipe node, rotate=90, anchor=base east]
     at (308.17, 791.98) {$\times 2$\,};
  \draw[-{>[ipe arrow small]}]
    (288, 696)
     -- (288, 800)
     -- (328, 800)
     -- (328, 704);
\end{tikzpicture}\endgroup \renewcommand{\baselinestretch}{1.5}}}
    \caption{Benchmark phase difference prediction architecture, using causal 2D convolutions over time and frequency dimensions \cite{masuyamaOnlinePhaseReconstruction2023}.} \vspace{-1em}
    \label{fig:masuyama_arch}
\end{figure}
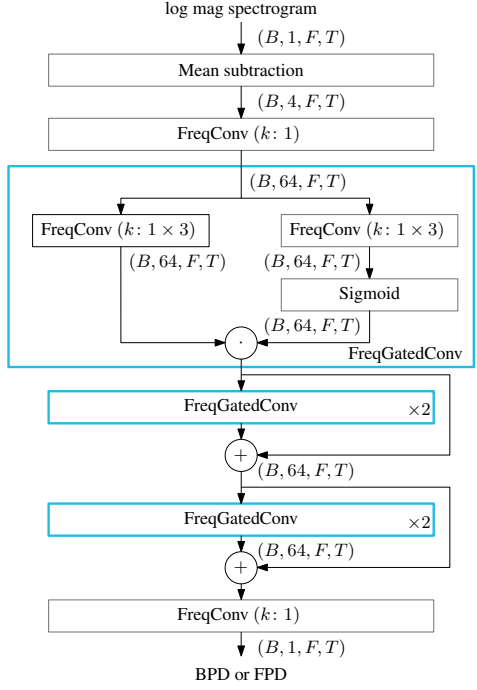

We adapt the Vocos backbone to predict \gls{bpd} and \gls{fpd}, by altering the first Conv1D layer to take a 513-channel log-magnitude spectrogram, and the 1026-channel output is split into \gls{bpd} and \gls{fpd}. For comparison, we also design a Conv2D variant with 6 ConvNeXt blocks using (32, 64) channels instead of (512, 1536). This results in approximately $400\times$ fewer parameters. The phase difference prediction method proposed by Masuyama et al., whose architecture is shown in Fig.\,\ref{fig:masuyama_arch}, serves as a benchmark. To train these models, we use the linear wrapping-aware loss \cite{aiLowLatencyNeuralSpeech2024,duGANNecessaryMelSpectrogramBased2025}, given by

\begin{equation}
    \mathcal{L}_{\text{wa}}(\hat x, x) = \left\Vert \hat x - x - 2\pi \text{ round}\left(\frac{\hat x - x}{2\pi}\right) \right\Vert_1~. \label{eq:wrapping_aware}
\end{equation}

For evaluation, the predicted phase differences need to be converted into a phase spectrogram $\bm{\hat \phi}$. We do so by using the least-squares phase reconstruction method of Masuyama et al. \cite{masuyamaOnlinePhaseReconstruction2023}.

\section{Experimental Setup}

\subsection{Datasets}
\label{sec:datasets}

For our trainings, we used LibriTTS \cite{zenLibriTTSCorpusDerived2019} subsets (train-clean-100, train-clean-360, train-other-500), similar to BigVGAN and Vocos \cite{leeBigVGANUniversalNeural2023, siuzdakVocosClosingGap2024}. This dataset contains approximately 585\,h of speech data from 2456 speakers. We employed the random amplitude augmentation as proposed for Vocos, namely, scaling the audio so that the maximum amplitude is within $[-6, -1)$\,dBFS. For validation and testing, we used LibriTTS subsets dev-clean (40 speakers, 9\,h) and test-clean (39 speakers, 8.5\,h), respectively. In addition, we used 20 proprietary German utterances from two speakers for evaluation. All audio files were resampled to 22.05\,kHz. While reporting the results, we use mnemonics D1 and D2 for test-clean and the out-of-distribution German datasets, respectively.

\subsection{Losses and training strategy} We use the following losses in our vocoding experiments: 1) the multi-resolution mel spectrogram loss $\mathcal{L}_\text{mel}$, 2) the hinge adversarial loss $\mathcal{L}_\text{adv}$, and 3) the feature matching loss $\mathcal{L}_\text{fm}$ of DAC \cite{kumarHighFidelityAudioCompression2023}. The discriminators were trained using the hinge critic loss of DAC, symbolized as $\mathcal{L}_{\text{critic}}$. For training phase difference prediction models, we use the wrapping-aware loss \eqref{eq:wrapping_aware}. The training settings for both lines of experiments are summarized in Table\,\ref{tab:training_strat}. 

\begin{table}[!t]
    \centering
    \caption{Training strategy for neural vocoding and phase difference prediction experiments, largely based on \cite{kumarHighFidelityAudioCompression2023} and \cite{masuyamaOnlinePhaseReconstruction2023}.}
    \label{tab:training_strat}
    {
        \footnotesize
        \begin{tabular}{l c c}
        \toprule
        Parameter  & Vocoding & Phase Diff.\\
        \midrule
        Batch size $(B)$ & 16 & 32 \\
        Utterance length [samples]& 16384 & 16384 \\
        Generator Loss & $15 \mathcal{L}_{\text{mel}} + \mathcal{L}_{\text{adv}} + 2 \mathcal{L}_{\text{fm}}$ & $\mathcal{L}_{\text{wa}}$ \\
        Discriminator Loss & $\mathcal{L}_{\text{critic}}$ & N/A \\
        Optimizer(s) & AdamW & RAdam \\
        Opt. params & lr:\,$2\mathrm{e}{-4}$, $\beta$:\,(0.9, 0.999) & lr:\,$6\mathrm{e}{-4}$ \\
        LR scheduler(s) & cosine annealing & cosine annealing  \\
        Training duration [steps] & 2M & 400k \\
        \bottomrule
        \end{tabular} \vspace{-1em}
    }
\end{table}

\section{Results}

\subsection{Comparison of vocoders}\label{subsec:results_vocoders}

\textbf{Objective evaluation. }We report five objective metrics: the multi-resolution log-mel spectrogram loss $\mathcal{L}_{\text{mel}}$, SCOREQ using the ground-truth audio as reference \cite{raganoSCOREQSpeechQuality}, voiced-unvoiced F1 score, periodicity RMSE \cite{morrisonChunkedAutoregressiveGAN2022}, and \gls{lsc} \cite{masuyamaOnlinePhaseReconstruction2023}
\begin{equation} \label{eq:lsc}
    \mathrm{LSC}(\bm{\hat{\phi}}, \mathbf{A}) = 20 \log_{10} \left(
        \frac{\Vert \mathbf{A} - |\mathrm{STFT}(\mathrm{iSTFT}(\mathbf{A}e^{j \bm{\hat \phi}}))| \Vert_{\mathrm{Fro}}}{\Vert \mathbf{A} \Vert_{\mathrm{Fro}}}
    \right)~,
\end{equation}
where $\mathbf{A}$ and $\bm{\hat{\phi}}$ denote the ground-truth magnitude spectrogram and the predicted phase spectrogram (obtained by applying STFT to the predicted waveform), respectively, and $\Vert \cdot \Vert_{\mathrm{Fro}}$ denotes the Frobenius norm. The results are reported in Table~\ref{tab:results_obj}.


\begin{table*}[ht]
    \centering
    \caption{Objective evaluation results, comparing Vocos to time-domain vocoders. BigVGANv2 (official) is the official checkpoint, and Vocos (official\textsuperscript{\textdagger}) is obtained by retraining Vocos using its official training recipe, but with the input representation explained in Section\,\ref{ssec:input_representation}.}
    {   
        \footnotesize
        \setlength{\tabcolsep}{0.9em}
        \begin{tabular}{l c cc cc cc cc cc}
            \toprule
            Model                   & \# params. & \multicolumn{2}{c}{$\mathcal{L}_{\text{mel}}$\,(\da)} & \multicolumn{2}{c}{SCOREQ\,(\da)} & 
                                        \multicolumn{2}{c}{V/UV F1\,(\ua)} & \multicolumn{2}{c}{Periodicity RMSE\,(\da)} & \multicolumn{2}{c}{\gls{lsc} [dB]\,(\da)} \\
            \cmidrule(lr){3-4} \cmidrule(lr){5-6} \cmidrule(lr){7-8} \cmidrule(lr){9-10} \cmidrule(lr){11-12} 
                                    &          & D1 & D2 & D1 & D2 & D1 & D2 & D1 & D2 & D1 & D2\\
            \cmidrule(lr){3-4} \cmidrule(lr){5-6} \cmidrule(lr){7-8} \cmidrule(lr){9-10} \cmidrule(lr){11-12} 
            BigVGANv2 (official)                          & 112M  & $0.767$ & $0.722$ & $.092$ & $.130$ & $.967$ & $.958$ & $.0665$ & $.0635$ & $-18.94$ & $-22.01$ \\
            Vocos (official\textsuperscript{\textdagger}) & 13.5M & $1.389$ & $1.463$ & $.161$ & $.213$ & $.938$ & $.917$ & $.1261$ & $.1380$ & $-12.68$ & $-13.54$ \\
            \cmidrule(r){1-2} \cmidrule(lr){3-4} \cmidrule(lr){5-6} \cmidrule(lr){7-8} \cmidrule(lr){9-10} \cmidrule(lr){11-12} 
            BigVGANv2                                     & 112M  & $0.846$ & $0.815$ & $.100$ & $.149$ & $.964$ & $.955$ & $.0713$ & $.0666$ & $-18.06$ & $-20.32$ \\
            BigVGANv2-base                                & 14.4M & $1.172$ & $1.126$ & $.144$ & $.194$ & $.948$ & $.934$ & $.1119$ & $.1172$ & $-13.70$ & $-15.16$ \\
            Vocos (modified)                              & 13.5M & $1.308$ & $1.309$ & $.188$ & $.258$ & $.940$ & $.919$ & $.1262$ & $.1374$ & $-12.39$ & $-13.43$ \\
            \cmidrule(r){1-2} \cmidrule(lr){3-4} \cmidrule(lr){5-6} \cmidrule(lr){7-8} \cmidrule(lr){9-10} \cmidrule(lr){11-12} 
            Vocos-Mag                                     & 13.2M & $0.495$ & $0.552$ & $.073$ & $.147$ & $.987$ & $.987$ & $.0273$ & $.0269$ & $-29.18$ & $-31.04$\\
            Vocos-Phase                                   & 13.2M & $1.560$ & $1.634$ & $.495$ & $.614$ & $.921$ & $.901$ & $.1755$ & $.1777$ & $-10.39$ & $-11.20$ \\

            \bottomrule
        \end{tabular}
        \vspace{-1em}
    }
    \label{tab:results_obj}
\end{table*}

Even though retraining on \mbox{LibriTTS} and cutting down the number of parameters degrade the objective metrics slightly, BigVGANv2 outperforms Vocos. Compared to the official Vocos recipe, using the outlined state-of-the-art vocoder training recipe yields mixed results for Vocos: improving $\mathcal{L}_{\text{mel}}$, deteriorating SCOREQ, and having negligible effects on pitch metrics and \gls{lsc}. 
\vspace{0.5em}

\noindent\textbf{Subjective evaluation. }Furthermore, we conducted a \gls{mushra} test to assess the subjective quality, using webMUSHRA \cite{schoefflerWebMUSHRAComprehensiveFramework2018}. Sixteen participants rated 14 out of the 20 aforementioned German audio samples  (dataset D2). For the lower anchor, we used pseudo-inverted mel spectrograms and the phase estimates by the PGHI algorithm \cite{prusaNoniterativeMethodReconstruction2017}, as also done in a previous work \cite{govalkarComparisonRecentNeural2019}. The test stimuli are available at our accompanying website\footnote{\label{fn:website}\url{https://audiolabs-erlangen.de/resources/NLUI/2026-IWAENC-vocoder}} 

\begin{figure}[!t]
    \centering
    {\begingroup
\renewcommand{\baselinestretch}{1} \endlinechar=-1 \input{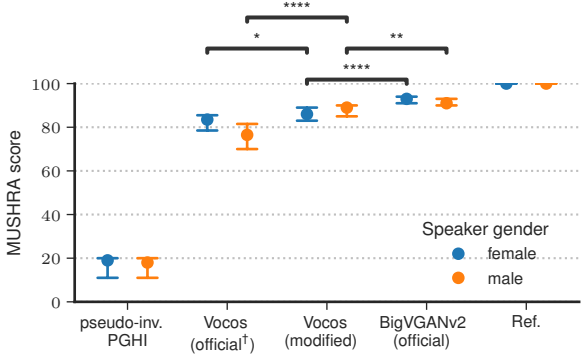}\endgroup \renewcommand{\baselinestretch}{1.5}} \vspace{-1.00em}
    \caption{Subjective listening test results (MUSHRA). The significance markers (`*': $p<0.05$, `**': $p<0.01$, and `****': $p<0.0001$) are based on two-sided Mann-Whitney-Wilcoxon tests with Holm-Bonferroni correction.} \vspace{-1.00em}
    \label{fig:subjective_test}
\end{figure}

Results, shown in Fig.\,\ref{fig:subjective_test}, corroborate that BigVGANv2 is significantly better than Vocos at reconstructing signals from band-limited mel spectrograms. Changes to the magnitude clamping and adopting a state-of-the-art training recipe yield a significant improvement for Vocos, but not enough to reach the performance of BigVGANv2. Interestingly, this outcome is only reflected by one of the objective metrics, namely $\mathcal{L}_{\text{mel}}$. The limited bandwidth of the underlying Wav2Vec2.0 representation might be rendering SCOREQ less sensitive to the differences between the two models.

\subsection{Learning isolated STFT components with Vocos}

Results from the previous section indicate that the gap between Vocos and BigVGANv2 is still present, even after controlling for confounding factors. To investigate the reasons for this gap, we trained two Vocos variants: 1) Vocos-Mag, which predicts only the log magnitude spectrogram $\mathbf{\hat{m}}$, and $\mathbf{\hat{p}}$ is set to the ground-truth phase; and 2) Vocos-Phase predicts the phase spectrogram $\mathbf{\hat{p}}$, and $\mathbf{\hat{m}}$ is set to the ground-truth log magnitudes. The results, reported in the bottom rows of Table~\ref{tab:results_obj}, highlight that Vocos-Mag outperforms almost any other experiment in this paper, while Vocos-Phase performs the worst. We draw a number of conclusions from these results:
\begin{itemize}
    \item The Vocos architecture can easily learn to predict a magnitude spectrogram that is compatible with the ground-truth phase. One dimensional convolutions seem to be sufficient for this task, calling into question modifications such as using a pseudo-inverse for recovering the linear-scale magnitude spectrograms \cite{liLearningNeuralVocoder2025}
    \item In contrast, predicting a phase spectrogram that is compatible with the ground-truth magnitudes is significantly more challenging, to the point that joint prediction of magnitude and phase spectrograms yields better results than predicting the phase spectrogram alone. While it is known that Vocos struggles with phase modeling, our findings show that the freedom to predict an `inconsistent' magnitude spectrogram (instead of being forced to predict the ground truth) is instrumental for Vocos' plausible waveform synthesis. This implies that approaches such as regularizing the predicted STFT coefficients to be closer to the ground-truth ones \cite{duAPNet2HighQualityHighEfficiency2024,duGANNecessaryMelSpectrogramBased2025, liLearningNeuralVocoder2025} may be counterproductive.
\end{itemize}

\subsection{Phase difference prediction with the Vocos backbone}

To isolate the source of Vocos' phase errors, we investigate if its backbone can predict phase differences from ground-truth log-magnitude spectrograms. This is important because, according to the signal model, phase differences are a precursor for phase reconstruction and can be estimated without autoregression. So, if the Vocos backbone is not effective for this task, it would indicate that the issue is not just with autoregression, but also with the inductive biases of the backbone. 

\begin{table}[!t]
    \centering
    \setlength{\tabcolsep}{0.28em}
    \caption{Impact of architectural choices on phase difference prediction performance.}
    {
        \footnotesize
        \begin{tabular}{l c c c c}
            \toprule
            Model & \# params. & \# FLOPs/s & $\mathcal{L}_{\text{wa}}$\,(\da) & \gls{lsc} [dB]\,(\da) \\
            \midrule
            Vocos backbone (Conv1D) & 15.0M & 3.0B & $.646$ & $-20.24$ \\
            Vocos backbone (Conv2D) & 37.1K & 3.5B & $.112$ & $-29.56$ \\
            Benchmark model \cite{masuyamaOnlinePhaseReconstruction2023} & 247K & 21.8B & $.238$ & $-26.12$ \\
            \bottomrule
        \end{tabular}
    } \vspace{-1em}
    \label{tab:phase_diff}
\end{table}

The results, reported in Table~\ref{tab:phase_diff}, reveal that the Vocos backbone with Conv1D layers is not effective for predicting phase differences. In contrast, switching to Conv2D layers substantially improves the performance while preserving the computational complexity and requiring much fewer learnable parameters. It even outperforms the benchmark method of Masuyama et al. \cite{masuyamaOnlinePhaseReconstruction2023}. This suggests that the inductive biases of the Vocos architecture, in particular the use of 1D convolutions, bottleneck its ability to model the time-frequency structure of speech signals, which is crucial for phase reconstruction. 

However, switching  to 2D convolutions for vocoding poses challenges. Frequency bins would become a spatial dimension, and thus narrow receptive fields of 2D convolutions could not learn harmonic structures spanning a wider frequency range. Therefore, future research should focus on inductive biases that better model the time-frequency structure of speech signals, without relying on ad-hoc remedies such as harmonic priors \cite{yoneyamaWavehaxAliasingFreeNeural2025} that rely on pitch information and limit applicability to inputs with local time-frequency structure.

\section{Conclusion}
In this paper, we investigated the limitations of the Vocos architecture. We showed that using a bandlimited mel spectrogram as input is an informative benchmark for comparing vocoders, accentuating the gap between time-domain and time-frequency domain vocoders. We found that even after controlling for confounding factors, Vocos still lags behind BigVGAN, a state-of-the-art time-domain vocoder. Then, to investigate the reasons for this gap, we trained Vocos variants with oracle knowledge of either the magnitude or phase spectrograms. Our results showed that the inconsistencies in Vocos-predicted magnitude spectrograms are \say{not a bug, but a feature}, as the freedom to predict an inconsistent magnitude spectrogram is instrumental for compensating errors in the phase. We conclude that a promising research direction for improving the phase reconstruction capabilities of Vocos is stronger inductive biases for modeling the time-frequency structure of speech signals.

\clearpage

\renewcommand{\refname}{REFERENCES}
\printbibliography[heading=bibnumbered]

\end{document}